\documentclass[aps,prb,twocolumn,amsmath,amssymb,superscriptaddress,floatfix]{revtex4}
\usepackage{graphicx}
\bibstyle{apsrev.bib}
\newcommand{\moy}[1]{\overline{\langle{#1}\rangle}}

\begin{document}

\title{Two-dimensional Ising model with self-dual biaxially correlated disorder}

\author{Farkas \'A. Bagam\'ery}
 \affiliation{Institute of Theoretical Physics,
Szeged University, H-6720 Szeged, Hungary}
 \affiliation{Laboratoire de Physique des Mat\'eriaux, Universit\'e Henri 
Poincar\'e (Nancy 1), BP~239,\\
F-54506 Vand\oe uvre l\`es Nancy Cedex, France}
\author{Lo\"{\i}c Turban}
 \email{turban@lpm.u-nancy.fr}
 \affiliation{Laboratoire de Physique des Mat\'eriaux, Universit\'e Henri 
Poincar\'e (Nancy 1), BP~239,\\
F-54506 Vand\oe uvre l\`es Nancy Cedex, France}
\author{Ferenc Igl\'oi}%
 \email{igloi@szfki.hu}
 \affiliation{Research Institute for Solid State Physics and Optics,
H-1525 Budapest, P.O.Box 49, Hungary}
 \affiliation{Institute of Theoretical Physics,
Szeged University, H-6720 Szeged, Hungary}

\date{\today}

\begin{abstract}
  We consider the Ising model on the square lattice with biaxially
  correlated random ferromagnetic couplings, the critical point of
  which is fixed by self-duality. The disorder, which has a
  correlator $G(r) \sim r^{-1}$, represents a relevant perturbation
  according to the extended Harris criterion.  Critical properties of
  the system are studied by large scale Monte Carlo simulations. The
  correlation length critical exponent $\nu=2.005(5)$ corresponds to
  that expected in a system with isotropic correlated long-range
  disorder, whereas the scaling dimension of the magnetization
  density $x_m=\beta/\nu=0.1294(7)$ is somewhat larger than in the
  pure system.  Conformal properties of the magnetization and energy
  density profiles are also examined numerically.
\end{abstract}

\pacs{}

\maketitle

\section{Introduction}

The presence of quenched disorder of different types is an inevitable
feature of real materials. The effect of randomness on the physical
properties of systems is often dramatic, in particular in the vicinity
of singular points, such as for classical phase transitions at finite
temperature or for quantum phase transitions at $T=0$.  In a random
system the set of critical exponents associated with a phase transition can be
completely different from that in the nonrandom system, which means
that disorder can be a relevant perturbation in the sense of the Harris
criterion.\cite{harris74} In most of the studies of disordered systems
the parameters (couplings, longitudinal or transverse fields, etc.)
are expected to be independent and identically distributed random
variables among which there is no correlation.

There are, however, systems in which the random variables exhibit some
kind of correlations, the effect of which can even change the random
universality class of the phase transition.  Examples of systems with
strongly correlated disorder are random quantum magnets at $T=0$, in
which the disorder is strictly correlated in the temporal direction.
Indeed the critical properties of a such type of random systems are quite
unusual, for a review see Ref.~\onlinecite{review}.  In the classical counterpart
of these quantum systems, the disorder is strictly correlated in one
direction, which leads to models introduced and partially exactly
solved by McCoy and Wu.\cite{mccoy68} In a further generalization,
the disorder can be strictly correlated in a $d'<d$ dimensional
subspace, where $d$ is the dimension of the
system.\cite{dorogovtsev80,boyanovsky82,lee92,blavatska03,sknepnek04}

Correlations between random variables can also be isotropic. One
example is presented in Ref.~\onlinecite{thurston94} in which the
experimental observations are explained\cite{altarelli95} with the
assumption of the presence of randomly oriented dislocation lines in
the sample.  Also in some random quantum magnets\cite{rieger99} and in
the non-Fermi liquid behavior of $f$-electron
compounds\cite{castroneto98} the disorder is expected to be
isotropically correlated due to long-ranged Ruderman-Kittel-Kasuya-Yoshida (RKKY) interactions. In a
coarse-grained picture, the disorder in some coupling strength $\Delta
K({\bf r})=K({\bf r})-\overline{K({\bf r})}$ is characterized by its
correlator, $G({\bf r}-{\bf r'})=\overline{\Delta K({\bf r}) \Delta
  K({\bf r'})}$, which is expected to be isotropic, $G({\rm
  r})=G(|{\rm r}|)$, and to decay algebraically for large arguments:
$G(r) \sim r^{-a}$.  For example, randomly oriented straight
dislocation lines are associated with a correlator exponent $a=d-1$.

The influence of Gaussian disorder with power-law correlations on the
critical behavior of the $m$-component vector spin model has been
studied in Ref.~\onlinecite{weinrib83} by field-theoretical methods
using a double expansion in $\epsilon=4-d$ and $\delta=4-a$.
Extending the Harris criterion, correlated disorder with $a\geq d$ is
found to be effectively short range and thus relevant only when the
pure system has a positive specific heat exponent.\cite{harris74} When
$a<d$ and $m=1$ (Ising model) the correlated disorder is shown to be a
relevant perturbation at the short-range-disorder fixed point,
provided the correlator decays sufficiently slowly, i.e., when
\begin{equation}
a < \frac{2}{\nu_{\rm short}}\,,
\label{harris}
\end{equation}
where $\nu_{\rm short}$ is the correlation length exponent of the system 
with short-range disorder.  At the stable long-range-disorder fixed point the
correlation length critical exponent is found to be given by
\begin{equation}
\nu=2/a,\quad 2/a > \nu_{\rm short}\,.
\label{nu_ft}
\end{equation}
This result is argued to be exact and thus the extended Harris
criterion in Eq.~(\ref{harris}) is marginal when the long-range nature
of the correlations is relevant. On the other hand the magnetic
exponents, such as $x_m=\beta/\nu$, depend on both $a$ and $d$.

Recently,\cite{prudnikov00} the validity of the scaling
relation~(\ref{nu_ft}) has been questioned on the basis of a direct
renormalization analysis of the scaling functions for the three-dimensional (3D) system
in the two-loop approximation, using a Pade-Borel summation technique
to evaluate the critical exponents for different values of $a$.

Numerical and field-theoretical calculations on different types of
systems with correlated disorder (percolation,\cite{weinrib84} 3D
Ising model,\cite{ballesteros99} polymers,\cite{blavatska01} quantum
Ising model,\cite{rieger99} etc.) seem to be in overall agreement with
the prediction of Eq.~(\ref{nu_ft}). 

Here we should make, however, two
remarks. First, in the case of the Ising model with randomly distributed 
dislocation lines, oriented along the lattice
axes,\cite{ballesteros99} for which rotational symmetry is broken,
 the critical behavior was found to be the same as for the model 
 with Gaussian correlated disorder when $a=2$. This shows that 
 the details of the model of correlated disorder are seemingly irrelevant,
 the critical behavior being essentially governed by the decay of 
 the disorder correlations.
Our second remark
concerns the fact that, in systems with correlated disorder, the
location of the critical point is generally not known exactly, which
introduces a limitation on the accuracy of the numerical calculations.

In the present paper, our aim is to increase the numerical accuracy of
the calculation in a statistical mechanical system with correlated
disorder and address also such questions (c.f. critical point density
profiles, validity of conformal invariance, etc.) which have not been
considered previously. For this purpose, we consider the
two-dimensional (2D) square lattice Ising model and introduce a type
of biaxially correlated disorder which preserves self-duality, thus
the critical point is exactly known. Our model can be considered as a
2D self-dual version of the random dislocation model of Ballesteros
and Parisi.\cite{ballesteros99} In the vicinity of the critical point
the properties of the system are explored at the scale of the
(diverging) correlation length.  A correlator exponent, $a=1$, can be
deduced from the integral of the disorder correlator inside a large
square of linear size, $L$. Thus at a coarse-grained level our model
can be compared to a model with Gaussian correlated disorder for the
same value of $a$.

For the 2D Ising model uncorrelated disorder coupled to the local
energy density (such as dilution or random ferromagnetic couplings) is
a marginally irrelevant perturbation,\cite{dotsenko83,selke94} hence
critical singularities of the pure system are supplemented by
logarithmic corrections. In particular the correlation length exponent
$\nu_{\rm short}$ keeps its unperturbed value $\nu_0=1$ . As a
consequence from Eq.~(\ref{harris}) the biaxially correlated disorder
of our model is expected to be relevant. In order to explore the
critical properties of the system and to check the validity of
Eq.~(\ref{nu_ft}) we perform large scale Monte Carlo simulations.  In
particular the critical exponents are deduced from the finite-size
scaling behavior of various magnetic and thermal quantities at the
critical temperature, which is exactly known. We also determine the
critical profiles of the magnetization and energy densities in order
to check whether the forms obtained for the profiles through conformal
methods remain valid in the random system after averaging over the
disorder.

The structure of the paper is the following. The model, its
self-duality, and the relevance-irrelevance criterion are described in
Sec.~\ref{sec:model}. Finite-size scaling calculations of the critical
exponents are presented in Sec.~\ref{sec:fss} and the critical profiles
are studied by conformal methods in Sec.~\ref{sec:profil}. Our results
are discussed in Sec.~\ref{sec:discus}.

\section{Model of self-dual biaxially correlated disorder}
\label{sec:model}

\subsection{Random-bond Ising models}

We study the spin $1/2$ $2D$ Ising model on a square lattice with Hamiltonian
\begin{equation}
-\beta{\cal H}=\sum_{i=1}^L\sum_{j=1}^L\left(K^x_{ij}\sigma_{i,j}\sigma_{i+1,j}
+K^y_{ij}\sigma_{i,j}\sigma_{i,j+1}\right)\,,
\label{hamiltonian}
\end{equation}
where $\beta=1/k_BT$. The Ising spin $\sigma_{i,j}=\pm 1$ is
associated with the site $(i,j)$, located at the intersection between
column $i$ and line $j$.  $K^x_{ij}$ and $K^y_{ij}$ denote the random
couplings between $\sigma_{i,j}$ and its first-neighbors
$\sigma_{i+1,j}$ in the horizontal direction and $\sigma_{i,j+1}$ in
the vertical direction, respectively. In the following we use the
parametrization
\begin{equation}
e^{2K^{\tau}_{ij}}-1=2^{1/2+u^{\tau}_{ij}},\;
\label{u}
\end{equation}
with $\tau=x,y$. Under duality transformation\cite{baxter} the
horizontal and vertical directions are exchanged and we obtain a
simple relation between the $u^{x}_{ij}$ and $u^{y}_{ij}$ variables
\begin{equation}
u^x_{ij}\longrightarrow u^{y*}_{ij}=-u^x_{ij}  \,,\qquad
u^y_{ij}\longrightarrow u^{x*}_{ij}=-u^y_{ij}  \,.
\label{uij2}
\end{equation}
where the superscript, $^*$, is used to denote dual variables.

With different distributions of $u^{\tau}_{ij}$ and thus $K^{\tau}_{ij}$ different models
are defined, which we list below.

\subsubsection{Nonrandom model}

The nonrandom model with $u^{x}_{ij}=u^{x}$ and $u^{y}_{ij}=u^{y}$
has its critical point at $u^{x}=-u^{y}$, which follows from self
duality in Eq.~(\ref{uij2}).  For the isotropic model the critical
point is at $u^{x}=u^{y}=0$. The critical exponents are\cite{baxter}
$\nu=1$ and $x_m=\beta/\nu=1/8$.

\subsubsection{Random model with uncorrelated disorder}

In this case the $u^{\tau}_{ij}$ parameters are independent and
identically distributed (iid) random variables, which are taken from a
distribution $P(u)$. At the critical point of the model the
distribution is symmetric,\cite{dom_kinz} $P(u)=P(-u)$, which is also
a consequence of self-duality, see Eq.~(\ref{uij2}). The strength of
disorder $\Delta$ is measured by $\Delta^2=\overline{u^2}$. In a
coarse-grained picture cells, of linear size L, are defined with
cell variables, $u'=L^{-d}\sum u_{ij}$, where the sum runs over $L^d$
sites, and $d=2$ is the dimension of the system.  In the
coarse-grained description we have $\Delta'=\Delta L^{-d/2}$.  To
decide about relevance-irrelevance of a perturbation according to
Harris\cite{harris74} one should consider the ratio $\Delta'/t$ at a
length-scale, $L = \xi \sim t^{-\nu}$, where $t$ measures the distance
from the critical point. For the 2D Ising model this analysis predicts
a marginal effect of the disorder, actually a marginally irrelevant one. 
 As a consequence
critical singularities of the nonrandom model are supplemented by
logarithmic corrections.\cite{dotsenko83,selke94}

\subsubsection{Random model with isotropically correlated disorder}

In this case the $u^{\tau}_{ij}$ parameters are random numbers which
are correlated at different sites and we have for large spatial
separation
\begin{equation}
\overline{u_{ij} u_{i'j'}} \sim \left[(i-i')^2+(j-j')^2\right]^{-a/2}\;.
\label{corr_a}
\end{equation}
The strength of disorder in the coarse-grained picture reads
as\cite{weinrib83} $\Delta_c'=\Delta_c L^{-\omega}$, with
$\omega=\mathrm{min}(d,a)/2$, thus long-range correlation in the disorder can
be relevant for $a<d$. Indeed for the 2D Ising model with $a<2$ this
type of perturbation is relevant, see Eq.~(\ref{harris}) and the
correlation length critical exponent is conjectured to be exactly
given by Eq.~(\ref{nu_ft}). For isotropic correlated disorder the
location of the critical point is not known exactly which makes it
difficult to analyze numerical results in the critical regime.

\subsubsection{McCoy-Wu model}

In the McCoy-Wu model\cite{mccoy68} the disorder is strictly
correlated in columns (or lines)
\begin{equation}
u_{ij}^x=b_i^x (=a_j^x), \quad u_{ij}^y=a_i^y (=b_j^y)\;,
\label{mw}
\end{equation}
where the parameters $b_i^x,~a_i^y$ ($a_j^x, b_j^y$) are iid random
variables. Now in the coarse-grained description with cells of size $L$, 
the disorder strength behaves as
$\Delta_{MW}'=\Delta_{MW} L^{-1/2}$. This type of disorder represents
a relevant perturbation. According to exact and conjecturedly exact
results, this model displays a strongly
anisotropic scaling behavior,\cite{fisher92} so that $\ln \xi_{\perp} \sim
\sqrt{\xi}$, where $\xi$ ($\xi_{\perp}$) is the correlation length
in the nontranslationally invariant (translationally invariant) direction.
The critical exponents\cite{fisher92} are given by: $\nu=2$ and
$x_m=(3-\sqrt{5})/4$.

\subsection{Biaxially correlated disorder}
 
%%%%%%%%%%%%%% FIG 1 %%%%%%%%%%%%%%%%%%%%%%%%%%%%

\begin{figure}[tbh]
\includegraphics[width=8.2cm]{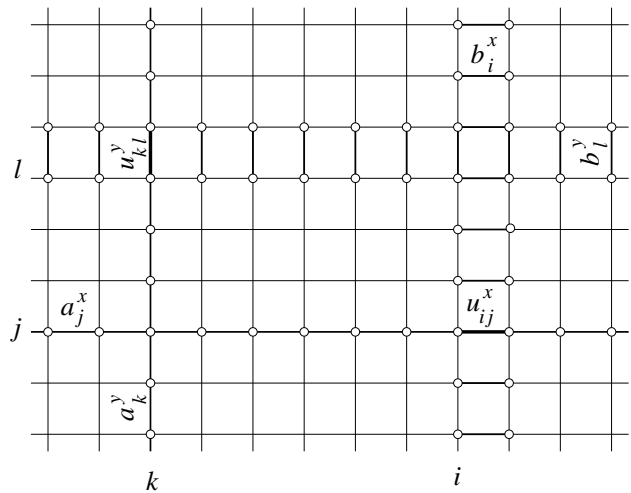}
\caption{Parameters entering in the definition of the correlated disorder.}
\label{fig1}
\end{figure}

%%%%%%%%%%%%%%%%%%%%%%%%%%%%%%%%%%%%%%%%%%%%%%

Here we introduce a type of correlated disorder which is

\begin{itemize}

\item Symmetric, with respect to the interchange of the $x$ and $y$ axis,

\item Has a duality symmetry and

\item Represents a 2D version of the random dislocation model.\cite{ballesteros99}

\end{itemize}

To be concrete we use the parametrization
\begin{equation}
u^x_{ij}=b_i^x+a_j^x\,,\qquad 
u^y_{ij}=a_i^y+b_j^y\,,
\label{uij1}
\end{equation}
(see Fig.~\ref{fig1}), which is the local average (for each bond) of
the couplings in two McCoy-Wu models where the disorder is along the
$x$ and the $y$ axis, respectively, see Eq.~(\ref{mw}).  Consequently
the strength of disorder is rescaled as in the McCoy-Wu model,
$\Delta'_b=\Delta_{b} L^{-1/2}$, but the system has isotropic scaling,
as far as the asymptotic behavior of the average correlations is
considered. This type of perturbation is
relevant (see Sec.\ref{relevance}) and has the same scaling exponent
as the isotropically correlated random perturbation with the
correlation parameter $a=1$ in Eq.~(\ref{corr_a}). We expect that the 
critical behavior is dictated by the averaged decay of the disorder correlations.
As for the 3D random dislocation model\cite{ballesteros99} where the 
dislocation lines are located along the lattice axes,
the breaking of rotational symmetry (before averaging over the quenched disorder) 
should not change the universality class. The present model, which has the 
same type of disorder correlations as  the Ballesteros-Parisi dislocation model, 
should also display the same critical behavior as the model with 
isotropically correlated disorder discussed in Ref.~\onlinecite{weinrib83}.

\begin{figure}[bht]
\includegraphics[width=\columnwidth]{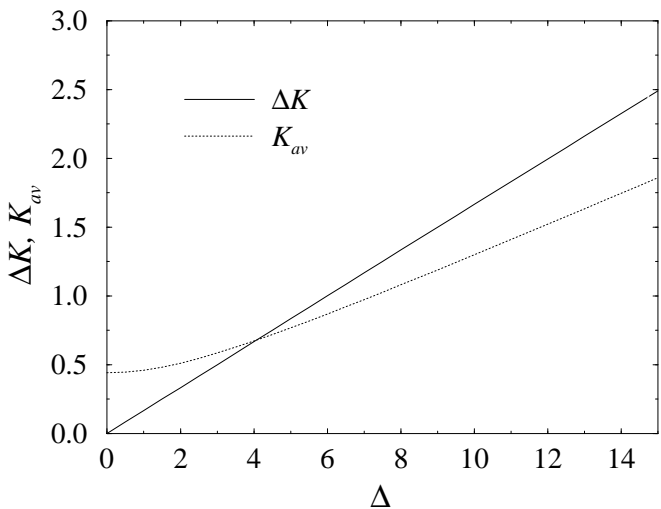}
\caption{Variation with $\Delta$ of the average value $K_{av}$ and the standard 
  deviation $\Delta K$ of the couplings. The standard deviation is
  almost linear in $\Delta$.}
\label{fig2}
\end{figure}

An important advantage of the parametrization used in Eq.~(\ref{uij1})
is that the couplings obey the duality relation in Eq.~(\ref{uij2}) and
thus the system is self-dual for an appropriate choice of the coupling
parameters $a_i^y,a_j^x,b_i^x$, and $b_j^y$.  Self-duality can be
realized in different ways, leading to a critical system in which both
lattice directions are statistically equivalent or not. Here we study
the geometry of disorder with the highest symmetry, i.e., with the
same symmetric and uniform probability density for all the random
variables
\begin{eqnarray}
P(s)&=&P(-s)=\left\{\begin{array}{ll}
\frac{1}{2\Delta}\quad & -\Delta\leq s\leq \Delta\\
\noalign{\vspace{3pt}}
0& |s|>\Delta
\end{array}\right.\nonumber\\
s&=&a_i^y,a_j^x,b_i^x,b_j^y\,.
\label{prodens}
\end{eqnarray}
The evolution of the average value $K_{av}$ and the standard deviation
$\Delta K$ of the coupling as a function of the half-width $\Delta$ of
the probability density is shown in Fig.~\ref{fig2}. The standard
deviation, with the expansion
\begin{equation}
\Delta K=.165764 \Delta + .000085 \Delta^3 + .000010 \Delta^5 +O(\Delta^7)\,,  
\label{deltak}
\end{equation}
is almost linear in $\Delta$. Evidently the relative strength of the disorder
$\Delta K/K_{av}$ is monotonously increasing with $\Delta$.

\subsection{Relevance-irrelevance criterion}
\label{relevance}

The stability of the fixed point governing the critical behavior of
the pure 2D system in the presence of correlated disorder can be
analyzed as by Harris.\cite{harris74,weinrib83} At the length scale
$L$, the sum of the deviations from the average coupling grows
typically like $\Delta L^{3/2}$ since it contains $2L$ independent
random variables with a vanishing average, each variable being of
order $\Delta L$ due to the correlations along the $L$ lines and the
$L$ columns. Dividing by the number of couplings one defines a typical
deviation from the average temperature as
\begin{equation}
\delta t(L)\simeq\Delta\frac{L^{3/2}}{L^2}\simeq\Delta L^{-1/2}\,.
\label{deltat1}
\end{equation}
%
%%%%%%%%%%%% FIG 2 %%%%%%%%%%%%%%%%%%%%%%%%%%%%%%
%%%%%%%%%%%%%%%%%%%%%%%%%%%%%%%%%%%%%%%%%%%%%%
The relevance of the perturbation depends on the behavior of the ratio  
\begin{equation}
\frac{\delta t(\xi_0)}{t}\propto t^{-1+\nu_0/2}\,,\qquad \xi_0\propto t^{-\nu_0}\,,
\label{deltat2}
\end{equation}
when one approaches the critical point ($t\to0$). Here $t=|T-T_{\rm
  c}|/T_{\rm c}$ is the reduced temperature, $\xi_0$ the correlation
length, and $\nu_0$ the correlation length exponent of the unperturbed
system. The perturbation is relevant when this ratio diverges at the
critical point, i.e., when $\nu_0<2$. Thus the biaxial correlated
disorder is a relevant perturbation in the case of the 2D Ising
model for which $\nu_0=1$.

Alternatively the same conclusion can be drawn from the value of the
scaling dimension of $\Delta$ following from Eq.~(\ref{deltat1}).
Under a change of the length scale by a factor $b$, $\delta t$, which
is a thermal perturbation, has a scaling dimension $y_{t0}=1/\nu_0$ and
thus transforms as
\begin{equation}
\delta t'=\Delta' L'^{-1/2}=b^{1/\nu_0}\delta t=b^{1/\nu_0}\Delta L^{-1/2}\,.
\label{deltat3}
\end{equation}
With $L'=L/b$ one obtains 
\begin{equation}
\Delta'=b^{y_{\!\Delta}}\Delta\,,\qquad y_{\Delta}=\frac{2-\nu_0}{2\nu_0}\,,
\label{ydelta}
\end{equation}
so that, in agreement with the previous result, one finds that the
strength of the disorder increases under rescaling when $\nu_0<2$.

Here we remind that the biaxial correlated disorder, which is
considered in this paper, can be represented by an effective
correlator exponent $a=1$ Indeed, the stability limit of the fixed
point for Gaussian correlated disorder in Eq.~(\ref{harris}) with
$a=1$ is the same as the one obtained here for biaxial correlations.
As mentioned in the Introduction, this result follows from the
comparison of the integral of the disorder correlator at a length
scale $L$ for both models.  The possible universality of the
critical properties of the system with respect to the form of
correlated disorder, in particular the validity of the result
predicted for the correlation length critical exponent in
Eq.~(\ref{nu_ft}) will be studied in the following sections.

\section{Finite-size scaling at the critical point}
\label{sec:fss}

\subsection{Monte Carlo technique}
We study the finite-size scaling behavior of different magnetic and
thermal quantities at the critical point. We work on a square-shaped system
with size $L$ and we use periodic boundary conditions in both
directions.

In order to limit the effects of the critical slowing down, the Monte
Carlo simulations are performed using the Wolff cluster
algorithm.\cite{wolff88} In order to estimate the equilibration time
$\tau_{\rm e}$ measured in cluster flips units, we compared the
evolution of the energy and the magnetization towards equilibrium
starting either from a random initial state or from an ordered initial
state. We have also studied the autocorrelation time for the energy
and the magnetization at equilibrium, in order to evaluate the
statistical errors on the time averages.

The time averages are calculated during a time $\tau=10^5$ to $2.5\times
10^5$ Monte Carlo steps (MCS) after a waiting time of the order of
$10\tau_{\rm e}$ ($10^3$ to $5\times 10^4$ MCS). The disorder averages are
taken over $n_{\rm s}\approx 10^4$ samples. These values were chosen
in order to obtain a precision close to $1\%$ for the different
quantities studied.

The simulations are performed in a vanishing external field and we use
the fluctuation-dissipation relations to evaluate the specific heat
and the magnetic susceptibility. The averaged value of a quantity $X$
is the result of a time average for a sample $s$ followed by an
average over $n_{\rm s}$ realizations of the disorder so that:
\begin{equation}
\moy{X}=\frac{1}{n_{\rm s}}\sum_{s=1}^{n_{\rm s}}\langle X^{(s)}\rangle\,,\qquad
\langle X^{(s)}\rangle=\frac{1}{\tau}\sum_{i=1}^{\tau}X_i^{(s)}\,.
\label{aver}
\end{equation}
The total magnetization $M$ and the total energy $E$ are defined as
\begin{equation}
M=\sum_{i,j=1}^{L}\sigma_{i,j}\,,\qquad
E=-\sum_{i,j=1}^{L}\left(\sigma_{i,j}\sigma_{i+1,j}+ 
\sigma_{i,j}\sigma_{i,j+1}\right)\,.
\label{me}
\end{equation}
The quantities studied are the following:
\begin{itemize}
\item The moments of the magnetization per site 
\begin{equation}
m_p=\frac{\moy{|M|^p}}{L^{2p}}\,,\qquad m_1=m\,.
\label{mp}
\end{equation}
\item The susceptibility per site
\begin{equation}
\chi=\frac{\moy{\Delta M^2}}{L^2}\,,\qquad \Delta M^2=(M-\langle M\rangle)^2\,.
\label{chi}
\end{equation}
%

%%%%%%%%%%% FIG 3  %%%%%%%%%%%%%%%%%%%%%%%%%%%%%%%

\begin{figure}[bht]
\includegraphics[width=\columnwidth]{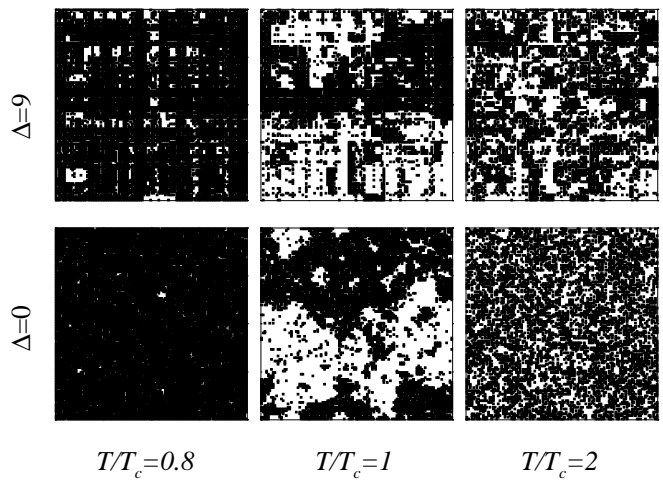}
\caption{Snapshots of equilibrium spin configurations for the random system ($\Delta=9$)
  and the pure system ($\Delta=0$) in the ordered phase ($T=0.8T_{\rm
    c}$), at the critical point and in the paramagnetic phase
  ($T=2T_{\rm c}$).}
\label{fig3}
\end{figure}

%%%%%%%%%%%%%%%%%%%%%%%%%%%%%%%%%%%%%%%%%%%%%%

%%%%%%%%%%%  FIG 4    %%%%%%%%%%%%%%%%%%%%%%%%%%%%%

\begin{figure}[thb]
\includegraphics[width=\columnwidth]{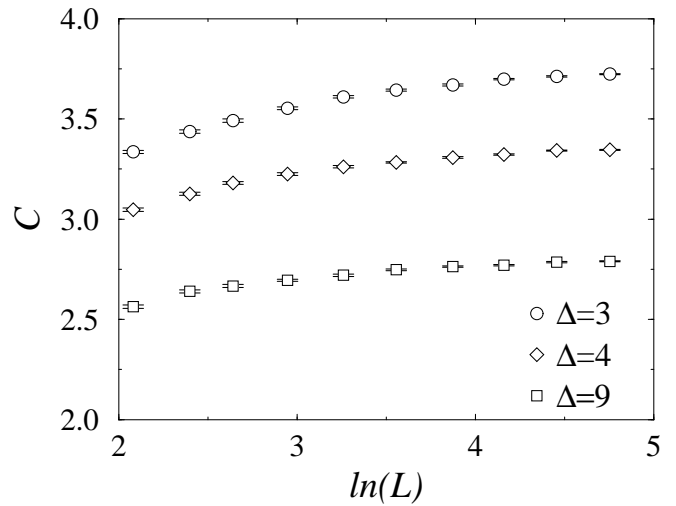}
\caption{Semilogarithmic plot of the finite-size behavior of the specific heat per spin at the critical point. The specific heat exponent $\alpha$ is negative and the behavior of $C$ is dominated by its regular contribution for large values of $L$.}
\label{fig4}
\end{figure}

%%%%%%%%%%%%%%%%%%%%%%%%%%%%%%%%%%%%%%%%%%%%%%

\item The specific heat per site
\begin{equation}
C=\frac{\moy{\Delta E^2}}{L^2}\,,\qquad \Delta E^2=(E-\langle E\rangle)^2\,.
\label{c}
\end{equation}
\item The temperature derivative of the logarithm of the susceptibility:
\begin{equation}
\Psi=\frac{\moy{\Delta E\,\Delta M^2}}{\moy{\Delta M^2}}\,,\qquad 
\Delta E=E-\langle E\rangle\,.
\label{psi}
\end{equation}
\end{itemize}

\subsection{Simulations results}
In Fig.~\ref{fig3} equilibrium spin configurations for the system with
biaxially correlated disorder at $\Delta=9$ are compared to the
corresponding pure system configurations in the low-temperature phase,
at the critical point and in the high-temperature phase for a system
with size $116\times116$ with periodic boundary conditions. In the
off-critical systems, spin clusters are larger for the disordered
system whereas the long-range aspect is roughly the same for both
systems at the critical temperature.

\begin{figure}[thb]
  \includegraphics[width=\columnwidth]{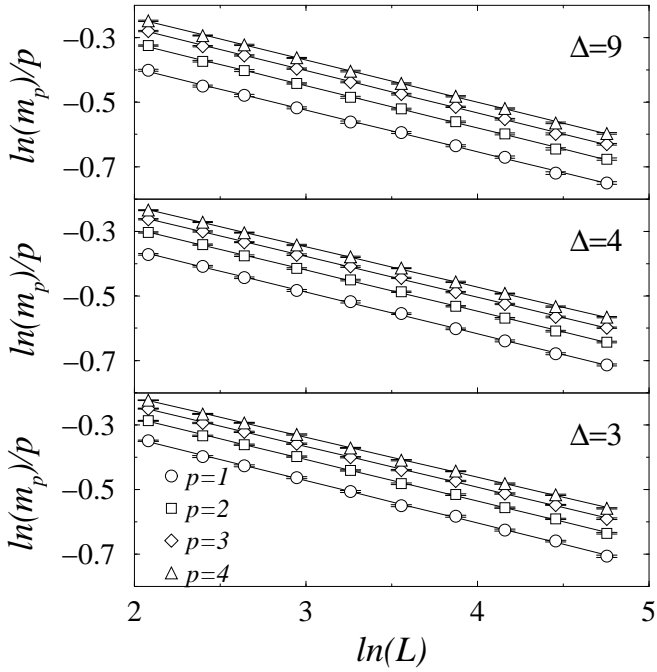}
\caption{Finite-size scaling of the moments of the magnetization density at the critical point. The lines correspond to linear fits of the data.}
\label{fig5}
\end{figure}

\begin{figure}[bht]
\includegraphics[width=8.2cm]{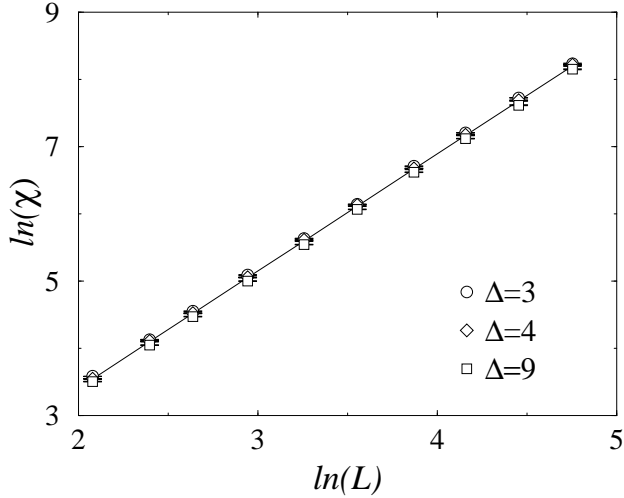}
\caption{Finite-size scaling of the magnetic susceptibility per spin at the critical point. The influence of $\Delta$ is quite small and the line gives the linear fit of all the data.}
\label{fig6}
\end{figure}

\begin{figure}[thb]
\includegraphics[width=\columnwidth]{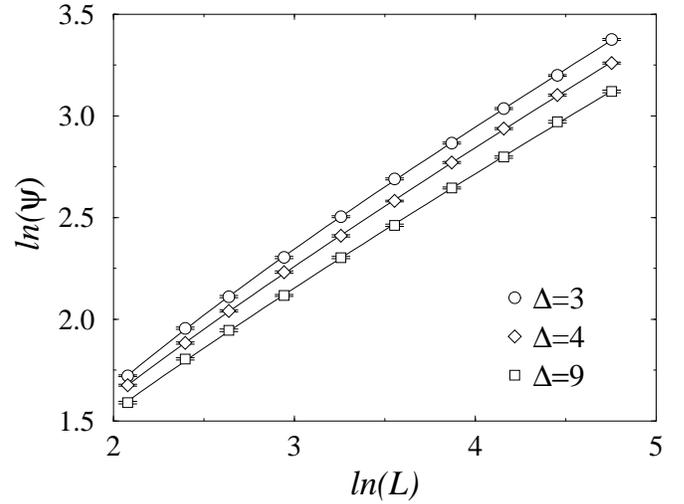}
\caption{Finite-size scaling of $\Psi$, temperature derivative of the magnetic susceptibility at the critical point. The lines are the nonlinear fits of the data as explained in the text.}
\label{fig7}
\end{figure}

The finite-size behaviors at the critical point of the specific heat
$C$, the moments of the magnetization density $m_p$, the
susceptibility $\chi$, and the temperature derivative of $\ln\chi$,
$\Psi$, are shown in Figs.~\ref{fig4}--\ref{fig7} for three values of
the disorder amplitude, $\Delta=3$, 4 and 9, and for ten sizes ranging
from $L=8$ to $L=116$.

%%%%%%%%%%% FIGS 5--6  %%%%%%%%%%%%%%%%%%%%%%%%%%%%%

%%%%%%%%%%%%%%%%%%%%%%%%%%%%%%%%%%%%%%%%%%%%%%

%%%%%%%%%%% FIGS 7 %%%%%%%%%%%%%%%%%%%%%%%%%%%%%

%%%%%%%%%%%%%%%%%%%%%%%%%%%%%%%%%%%%%%%%%%%%%%

%%%%%%%%%%% FIGS 8 %%%%%%%%%%%%%%%%%%%%%%%%%%%%%%%

\begin{figure}[bht]
  \includegraphics[width=\columnwidth]{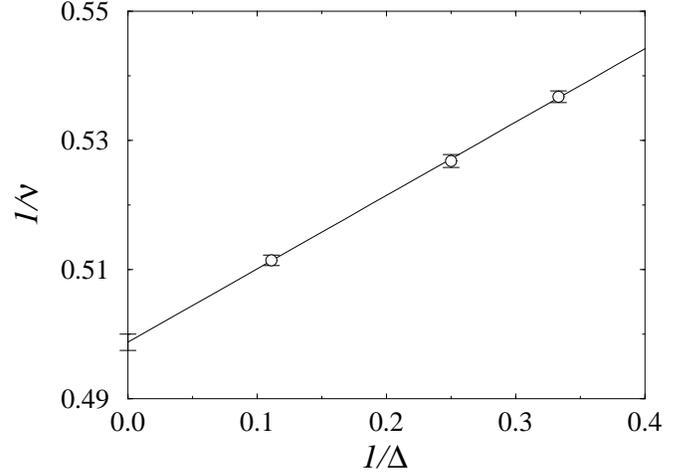}
\caption{Linear extrapolation of $1/\nu$ to infinite disorder strength.}
\label{fig8}
\end{figure}

%%%%%%%%%%%%%%%%%%%%%%%%%%%%%%%%%%%%%%%%%%%%%%

%%%%%%%%%%% TABLE 1 %%%%%%%%%%%%%%%%%%%%%%%%%%%%%%

\begin{table}
\caption{Effective critical exponents deduced from the fits of the finite-size scaling data in Figs.~\ref{fig5}--\ref{fig7}. The last line gives the values of the effective correction-to-scaling exponent for $\Psi$ defined in Eq.~(\ref{nlf}).\label{table1}}
\vskip 0.2truecm
\begin{ruledtabular}
\begin{tabular}{llll}
&$\Delta=3$&$\Delta=4$&$\Delta=9$ 
\\
\hline
$\beta/\nu$&0.1312(13)&0.1292(10)&0.1294(14) \\
$x_m^{(2)}$&0.1283(10)&0.1279(8)&0.1309(11)\\
$x_m^{(3)}$&0.1259(8)&0.1264(8)&0.1310(9)\\
$x_m^{(4)}$&0.1238(7)&0.1249(7)&0.1305(8)\\
$\gamma/\nu$&1.7435(21)&1.7442(17)&1.7381(22)\\
$1/\nu$&0.5368(9)&0.5268(10)&0.5115(8)\\
$\omega$&0.737(7)&0.573(6)&0.569(7)\\
\end{tabular}
\end{ruledtabular}
\end{table}

%%%%%%%%%%%%%%%%%%%%%%%%%%%%%%%%%%%%%%%%%%%%%%

The specific heat saturates at large system size, which indicates that
the regular contribution dominates the singular one. The singular
contribution, behaving as $L^{\alpha/\nu}$, vanishes at large system
size, i.e., the specific heat exponent $\alpha$ is negative. This is
the reason why we studied the finite-size behavior of $\Psi$ which
diverges with the size of the system, thus allowing us to estimate the
thermal exponent $1/\nu$.

According to finite-size scaling theory, the following behaviors are
expected for a large critical system with size $L$:
\begin{eqnarray}
m_p &\propto&  L^{-px_m^{(p)}}\,,\qquad x_m^{(1)}=x_m=\beta/\nu\,,\nonumber\\
 \chi &\propto&  L^{\gamma/\nu}\,,\qquad \gamma/\nu=d-2x_m=2(1-x_m)\,,\nonumber\\
 \Psi &\propto&  L^{1/\nu}\,,\qquad 1/\nu=y_t\,.
\label{fss}
\end{eqnarray}
The log-log plots for $(m_p)^{1/p}$, $\chi$, and $\Psi$ display the
expected linear behavior, although with noticeable deviations at small
size and a slightly $\Delta$-dependent slope for $\Psi$. Thus the
magnetization and susceptibility exponents were simply obtained
through linear fits, whereas we used a nonlinear fit of the data for
$\Psi$
\begin{equation}
\ln(\Psi)=A+\frac{1}{\nu}\ln(L)+\ln\left(1+BL^{-\omega}\right)\,.
\label{nlf}
\end{equation}
The exponents are collected in Table~\ref{table1} for the three values
of the disorder strength. There is no significant variation of
$\beta/\nu$ and $\gamma/\nu$ with $\Delta$. Taking the average of the
exponents obtained for the three values of $\Delta$ leads to
\begin{equation}
\frac{\beta}{\nu}=x_m=0.1299(12)\,,\quad \frac{\gamma}{\nu}=1.7419(20)\,.
\label{betagamma}
\end{equation}
There is no clear evidence of multiscaling: The slow decay of
$x_m^{(p)}$ with increasing $p$ is probably due to a crossover effect
since it disappears at $\Delta=9$ when the disorder is sufficiently
strong.

The variation of $1/\nu$ with $\Delta$ is significant. As shown in
Fig.~\ref{fig8} it is linear in $1/\Delta$ and the extrapolation leads
to
\begin{equation}
\frac{1}{\nu}=y_t=0.4988(13)\,,
\label{nu}
\end{equation}
at infinite disorder strength.

\section{Critical profiles}
\label{sec:profil}

The analysis of Monte Carlo data for the critical profiles using the
tools of conformal theory provides an efficient method for the
determination of bulk and surface critical
exponents.\cite{chatelain99,res00,berche03}

%%%%%%%%%%% FIG 9  %%%%%%%%%%%%%%%%%%%%%%%%%%%%%%%

\begin{figure}[bht]
\includegraphics[width=\columnwidth]{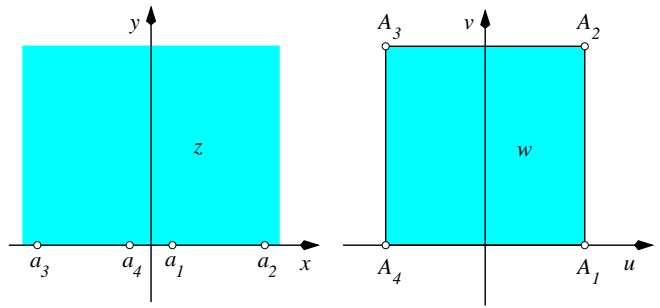}
\caption{Conformal transformation of the half-plane into a square.  The boundary conditions, free or fixed,  are the same on both systems.}
\label{fig9}
\end{figure}

%%%%%%%%%%%%%%%%%%%%%%%%%%%%%%%%%%%%%%%%%%%%%%

%%%%%%%%%%% FIG 10 %%%%%%%%%%%%%%%%%%%%%%%%%%%%%%%%

\begin{figure}[thb]
\includegraphics[width=\columnwidth]{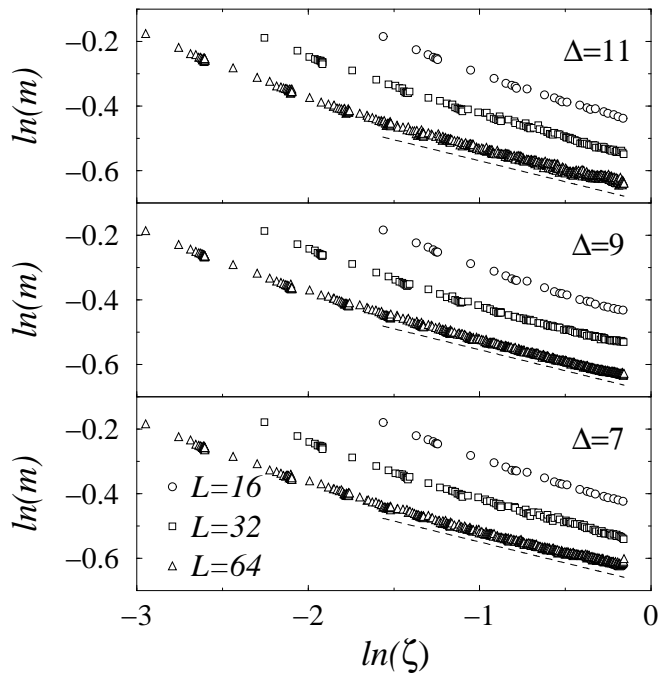}
\caption{Conformal profiles of the magnetization density for different system sizes $L$ and disorder strength $\Delta$. The dashed lines indicate the slope $-\beta/\nu$ expected from the finite-size scaling results.}
\label{fig10}
\end{figure}

%%%%%%%%%%%%%%%%%%%%%%%%%%%%%%%%%%%%%%%%%%%%%%

Using the replica trick for a system with biaxially correlated
disorder leads to an effective Hamiltonian with long-range
interactions between the replicas. Thus we do not expect our system to
be conformally invariant at its new fixed point as it seems to be the
case for uncorrelated disorder.\cite{chatelain99} Nonetheless, in this
section we assume that the critical profiles have a leading behavior
which is the same as for a conformally invariant system. In both cases
the form of the profiles is constrained by covariance under global
scale transformations and by the values of surface and bulk exponents.
We know of at least one example, the random Ising chain in a
transverse field, for which the critical profiles follow quite
accurately the conformal predictions,\cite{igloi98} although the
system cannot be conformally invariant since it displays a strongly
anisotropic scaling behavior.\cite{fisher92,young96}

\subsection{Conformal transformation of the densities}

%%%%%%%%%%% FIG 11  %%%%%%%%%%%%%%%%%%%%%%%%%%%%%%%

\begin{figure}[tbh]
  \vglue.2cm \includegraphics[width=\columnwidth]{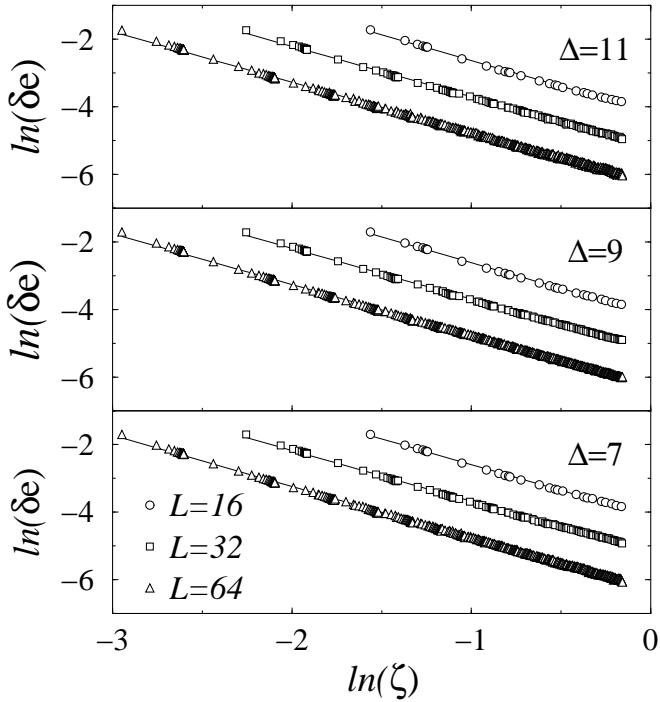}
\caption{Conformal profiles of the singular part of the energy density for different system sizes $L$ and disorder strength $\Delta$. The lines correspond to the linear fits of the data.}
\label{fig11}
\end{figure}

%%%%%%%%%%%%%%%%%%%%%%%%%%%%%%%%%%%%%%%%%%%%%%

On a semi-infinite critical system with fixed boundary conditions, the
form of the magnetization profile is fixed, up to a constant
amplitude, by translational, rotational, and global scale invariance.
It decays as a power of the distance $y$ from the surface
\cite{fisher78}
\begin{equation}
\moy{m(y)}_{\rm fixed}=A_m\,y^{-x_m}\,,\qquad x_m=\beta/\nu\,,
\label{my}
\end{equation}
where the exponent $x_m$ is the scaling dimension of the bulk magnetization density.

Let us consider the Schwarz-Christoffel  transformation of the half-plane $z=x+iy$ such that
\begin{equation}
w(z)=\frac{L}{2K(k)} F(z,k)\,,\quad z={\rm sn}\left(\frac{2Kw}{L}\right)\,,
\label{wz}
\end{equation}
where $w(z)=u+iv$ and
\begin{equation}
F(z,k)=\int_0^z \frac{dt}{\sqrt{(1-t^2)(1-k^2t^2)}}
\label{sk}
\end{equation}
is the elliptic integral of the first kind with modulus $k$.
$K(k)=F(1,k)$ is the complete elliptic integral of the first kind and $\mathrm{sn}$ is the Jacobian elliptic sine.

This conformal transformation maps the half-plane $y=\Im(z)\geq0$ onto
the interior of the square with $-L/2\leq u\leq L/2$, $0\leq v\leq L$
(see Fig.~\ref{fig9}) when the modulus $k$ verifies the relation
$K(k)/K(\sqrt{1-k^2})=1/2$ (see e.g. Ref.~\onlinecite{fuchs64}).

In the new geometry the magnetization density is given by
\begin{eqnarray}
\moy{m(w)}_{\rm fixed}&=&b(z)^{x_m}\moy{m(z)}_{\rm fixed}\,,\quad b(z)=| 
w'(z)|^{-1}\nonumber\\
w'(z)&=&\frac{L}{2K(k)\sqrt{(1-z^2)(1-k^2z^2)}}\,.
\label{ctm}
\end{eqnarray}
Finally, introducing the reduced variable
\begin{equation}
\zeta(w)=\frac{\Im[z(w)]}{\left|\sqrt{(1-z^2)(1-k^2z^2)}\right|}\,,
\label{zeta}
\end{equation}
the magnetization density varies as a power of $\zeta$
\begin{equation}
m=\moy{m(\zeta)}_{\rm fixed}\propto \zeta^{-x_m}\,.
\label{mzeta}
\end{equation}

The same behavior is obtained for the singular part of the energy density, for which
\begin{equation}
\moy{e(\zeta)}_{\rm sing}\propto \zeta^{-x_e}\,,\qquad x_e=2-1/\nu\,.
\label{ezeta}
\end{equation}
In this case one needs to eliminate the regular contribution to the
energy density. The regular part can be canceled by taking the
difference $\delta e$ between the energy density profiles obtained on
a system with either free boundary conditions $e_{\rm free}(w)$ or
fixed boundary conditions $e_{\rm fixed}(w)$ and then
 \begin{equation}
\delta e=\moy{e_{\rm free}(\zeta)}-\moy{e_{\rm fixed}(\zeta)}\propto 
\zeta^{-x_e}\,.
\label{dezeta}
\end{equation}
The amplitudes of the singular parts have opposite signs with the two
types of boundary conditions,\cite{karevski96,berche03} thus in the
difference the singular contribution is amplified and the bulk regular
contribution is eliminated.

%%%%%%%%%%% TABLE 2 %%%%%%%%%%%%%%%%%%%%%%%%%%%%%%

\begin{table}
\caption{Scaling dimension of the energy density $x_e$ deduced from the linear fits of the conformal profiles in Fig.~\ref{fig11}.\label{table2}}
\vskip 0.2truecm
\begin{ruledtabular}
\begin{tabular}{llll}
$L$&$\Delta=7$&$\Delta=9$&$\Delta=11$\\
\hline
16&1.5435(9)&1.5377(9)&1.5177(10)\\
32&1.5241(9)&1.5207(6)&1.5110(9)\\
64&1.5209(6)&1.5182(4)&1.5002(8)\\
\end{tabular}
\end{ruledtabular}
\end{table}

%%%%%%%%%%%%%%%%%%%%%%%%%%%%%%%%%%%%%%%%%%%%%%

%%%%%%%%%%% TABLE 3 %%%%%%%%%%%%%%%%%%%%%%%%%%%%%%

\begin{table*}
\caption{Critical exponents deduced from the scaling laws using the finite-size scaling results for $\beta/\nu$, $\gamma/\nu$ in Eq.~(\ref{betagamma}) and $1/\nu$ in Eq.~(\ref{nu}) as inputs. \label{table3}}
\vskip 0.2truecm
\begin{ruledtabular}
\begin{tabular}{llllllll}
$x_m$&$x_e$&$\alpha$&$\beta$&$\gamma$&$\delta$&$\nu$&$\eta$\\
\hline
0.1294(7)&1.5012(13)&-2.01(1)&0.2604(31)&3.492(13)&14.46(8)&2.005(5)&0.2588(14)\\
\end{tabular}
\end{ruledtabular}
\end{table*}

%%%%%%%%%%%%%%%%%%%%%%%%%%%%%%%%%%%%%%%%%%%%%%

\subsection{Monte Carlo simulations}
We work on a square lattice with either fixed or free boundary
conditions using the Swendsen-Wang cluster algorithm\cite{swendsen87}
which is more efficient than the Wolff algorithm with fixed boundary
conditions. The simulations were performed for three lattice sizes
($L=16$, $32$, $64$) and three values of the couplings ($\Delta=7$,
$9$, $11$). After a waiting time $\tau_{\rm e}=5\times10^3$ MCS, the time
averages are taken during a time $\tau=6\times 10^4$ MCS. The disorder
averages are taken over a number of samples ranging from $n_{\rm s}=2\times
10^4$ for the smaller systems to $n_{\rm s}=2\times 10^3$ for the larger
ones.

For the magnetization density profiles all the surface spins are fixed
in the $+1$ state.  The behavior of the singular part of the energy
density is obtained by taking the difference of Eq.~(\ref{dezeta})
between two systems with either free or fixed boundary conditions for
the {\it same} realizations of the disorder, which considerably
reduces the statistical noise.

The log-log plots for the magnetization and the energy densities are
shown in Figs.~\ref{fig10}--\ref{fig11} as functions of the reduced
conformal variable $\zeta$ defined in Eq.~(\ref{zeta}).

The magnetization density profiles show strong corrections to scaling
and due to the small values of $\zeta$ we were unable to find a
reliable nonlinear fit of our data. Nevertheless, the dashed lines
indicate that the asymptotic slopes are in agreement with the value of
$\beta/\nu$ obtained by finite-size scaling and given in
Eq.~(\ref{betagamma}).

On the contrary, the corrections are very weak for the energy density
profiles in Fig.~\ref{fig11} so that the slopes could be obtained with
a good accuracy through a linear fit of the data. The small deviations
from linearity occur close to the surface for small values of $\zeta$.
Only a few points are concerned and their weight in the fit is
negligible. The values of $x_e$ are collected in Table~\ref{table2}
for the different values of $L$ and $\Delta$. They are slowly
decreasing when $\Delta$ or $L$ increases.

\section{Discussion}
\label{sec:discus}

Our results are collected in Table~\ref{table3}. The scaling dimension
of the magnetization density $x_m$ is directly given by $\beta/\nu$
in Eq.~(\ref{betagamma}). Alternatively, it can estimated using
$\gamma/\nu$ together with the scaling law $\gamma/\nu=d-2x_m$, which
gives $x_m=0.1291(10)$. The value in Table~\ref{table3} corresponds to
the centre of the confidence interval which is common to the two
estimates. It is quite close to the value for the pure 2D
Ising model $x_m=1/8$. Thus the magnetization in the random and pure
systems have almost the same fractal dimensions. This explains why
both systems in Fig.~\ref{fig3} have roughly the same aspect at the
critical point. The value of $x_m$ was used to calculate
$\eta=2x_m+2-d=2x_m$ and $\delta=-1+d/x_m$.

The value of $\nu$ following from Eq.~(\ref{nu}) is such that,
according to Eq.~(\ref{ydelta}), the disorder is irrelevant at the new
fixed point, as required for its stability. Furthermore the numerical
estimate of $\nu$, within the error of the calculation corresponds to
the prediction in Eq.~(\ref{nu_ft}) with the effective correlator
exponent, $a=1$. The estimate of $\nu$ leads to the value of the
specific heat exponent through the Josephson scaling law
$\alpha=2-d\nu$. The correlation length diverges more strongly in the
random system with $\nu$ close to 2 than in the pure system with
$\nu=1$, this might explain the difference in the cluster sizes for
the off-critical systems in Fig.~\ref{fig3}.

The scaling dimension of the energy density $x_e$ is obtained using
the scaling law $x_e=d-1/\nu$. The values of $x_e$ deduced from the
conformal profiles, collected in Table~\ref{table2}, are close to this
estimate for large sizes and strong enough disorder.

To summarize we have studied the critical behavior of the square
lattice Ising model with biaxially correlated disorder by large scale
MC simulations. The different estimates of the critical exponents give
a consistent picture, in which $\nu$ seems to be the same as for
isotropic Gaussian correlated disorder, whereas $x_m$ is not very far
from the pure and uncorrelated random systems values. 

One important technical simplification of the model is its
self-duality property, which is used to locate exactly the critical
point.  The same type of parametrization, as given in
Eq.~(\ref{uij1}) can be used to preserve self-duality in the $q$-state
Potts model, for any value of $q$.\cite{Wu82} An investigation of
these models could provide further information about the nature of the
phase transition in disordered systems. For example, $q=1$ is
equivalent to the percolation problem,\cite{kasteleyn69} whereas the
phase transition in the pure system is of first order for
$q>4$.\cite{baxter73} In the latter problem, due to uncorrelated
disorder, the transition is softened into a second-order
one,\cite{aizenman89} with a correlation length exponent $\nu_{\rm
  short} \approx 1$.\cite{Cardy99} Consequently, according to the
extended Harris criterion in Eq.~(\ref{harris}), the phase transition
is expected to belong to a new and $q$-dependent universality class.
If the limiting behavior for large $q$ can be predicted, as for
uncorrelated disorder\cite{anglesdauriac03} remains the subject of a
separate study.

\begin{acknowledgments}
  We thank Christophe Chatelain for useful discussions
  and Bertrand Berche for giving us his tabulated
  values of $\zeta$.  F.\'A.B. thanks the Minist\`ere Fran\c{c}ais des
  Affaires \'Etrang\`eres for a research grant. This work has been
  supported by the Hungarian National Research Fund under Grant Nos.
  OTKA TO34183, TO37323, TO48721, MO45596, and M36803. Intensive
  simulations have been performed at CINES Montpellier under Project No.
  pnm2318.  The Laboratoire de Physique des Mat\'eriaux is Unit\'e
  Mixte de Recherche CNRS No. 7556.
\end{acknowledgments}

\end{document}